\documentclass[aps,prl,superscriptaddress,floatfix,showpacs,reprint]{revtex4-1}

\usepackage{verbatim}

\usepackage{amsmath}
\usepackage{amsfonts}
\usepackage{amssymb}
\usepackage{graphicx}
\usepackage{bm}
\usepackage{color}

\usepackage{epsf}

\usepackage{psfrag}

\newcommand{\veck}{{\bf k}}
\newcommand{\vecki}{{\bf k}_i}
\newcommand{\veckj}{{\bf k}_{J}}
\newcommand{\veckji}{{\bf k}_{J,i}}
\newcommand{\veckjone}{{\bf k}_{J,1}}
\newcommand{\veckjtwo}{{\bf k}_{J,2}}
\newcommand{\deins}[1]{{\rm d}#1\,}
\newcommand{\dzwei}[1]{{\rm d}^2#1\,}
\newcommand{\dk}{\dzwei{\veck}}

\newcommand{\dsigma}{\deins{\sigma}}
\newcommand{\dnu}{\deins{\nu}}
\newcommand{\dx}{\deins{x}}
\newcommand{\dyjetone}{\deins{y_{J,1}}}
\newcommand{\dyjettwo}{\deins{y_{J,2}}}
\newcommand{\dphij}{\deins{\phi_{J}}}
\newcommand{\dtwojets}{{\rm d}|\veckjone|\,{\rm d}|\veckjtwo|\,\dyjetone \dyjettwo}

\newcommand{\non}{\nonumber\\}
\newcommand{\asbar}{{\bar{\alpha}}_s}

\newcommand{\avgcos}{\langle \cos \varphi \rangle}
\newcommand{\avgcostwo}{\langle \cos 2 \varphi \rangle}
\newcommand{\avgcosthree}{\langle \cos 3 \varphi \rangle}

\newcommand{\init}{\text{init}}
\newcommand{\MOM}{\text{MOM}}
\newcommand{\MSbar}{\overline{\text{MS}}}
\newcommand{\BLM}{\text{BLM}}

\newcommand{\PS}{d({\rm P.\,S})}
\newcommand{\intPS}{\int \PS}

\begin{document}

\title
{Evidence for high-energy resummation effects in Mueller-Navelet jets at the LHC}

\author{B. Duclou\'e}
\affiliation{LPT, Universit{\'e} Paris-Sud, CNRS, 91405, Orsay, France}

\author{L. Szymanowski}
\affiliation{National Centre for Nuclear Research (NCBJ), Warsaw, Poland}

\author{S. Wallon}
\affiliation{LPT, Universit{\'e} Paris-Sud, CNRS, 91405, Orsay, France}
\affiliation{UPMC Univ. Paris 06, facult\'e de physique, 4 place Jussieu, 75252 Paris Cedex 05, France}

\begin{abstract}
\noindent
The study of the production of two forward jets with a large interval of rapidity
at hadron colliders was proposed by Mueller and Navelet as a possible test of 
the high energy dynamics of QCD. We analyze this process within a complete
next-to-leading logarithm framework, supplemented by the use of 
the Brodsky-Lepage-Mackenzie procedure extended to the perturbative Regge dynamics, to find the optimal renormalization scale.
This leads to a very good description of the recent CMS data at LHC for the 
azimuthal correlations of the jets.
\end{abstract}


\maketitle

\paragraph{Introduction.}

Many processes have been proposed as a way to probe the high energy dynamics of
QCD, described by the Balitsky-Fadin-Kuraev-Lipatov (BFKL)
approach~\cite{Fadin:1975cb,*Kuraev:1976ge,*Kuraev:1977fs,*Balitsky:1978ic}. 
Among the most promising ones is the 
production of two forward jets separated by a large interval of rapidity at 
hadron colliders, proposed by Mueller and Navelet~\cite{Mueller:1986ey}.
The purpose of the present work is to show that
the most recent LHC data extracted by the CMS collaboration
for the azimuthal correlations of these jets~\cite{CMS-PAS-FSQ-12-002}
are well described within this framework.

The description of this process involves two main building blocks: the jet 
vertex, which describes the transition from an incoming parton to a jet,
and the Green's function, which describes the pomeron exchange between the vertices.
The first results of a complete next-to-leading logarithmic (NLL) calculation, including the NLL corrections both to the Green's function~\cite{Fadin:1998py,*Ciafaloni:1998gs} and to
the jet vertex~\cite{Bartels:2001ge,*Bartels:2002yj}, showed that the NLL corrections to the jet vertex have a very large effect, leading to a lower cross section and a much larger azimuthal correlation~\cite{Colferai:2010wu}.
It was also observed that the results were very dependent on the choice
of the scales, especially the renormalization scale $\mu_R$ and the
factorization scale $\mu_F$. This has been confirmed in a more recent
study~\cite{Ducloue:2013hia}, where we used more realistic kinematic cuts. 
To reduce this dependency, we apply the physically motivated Brodsky-Lepage-Mackenzie (BLM)
procedure~\cite{Brodsky:1982gc} to fix the renormalization scale, 
as it was adapted 
to the resummed perturbation theory \`a la BFKL in Refs.~\cite{Brodsky:1998kn,*Brodsky:2002ka}.

\paragraph{Mueller-Navelet jets.}

The observables which are of interest are the differential cross section
 $\mathcal{C}_0$ 
\begin{equation}
  \mathcal{C}_0 = \frac{\dsigma}{\dtwojets}\,,
\end{equation}
where $\veckjone$, $\veckjtwo$ are the transverse momenta of the jets and
$y_{J,1}$, $y_{J,2}$ are their rapidities,
and the azimuthal correlations~\cite{DelDuca:1993mn,*Stirling:1994zs} of the jets 
\begin{equation}
  \frac{\mathcal{C}_n}{\mathcal{C}_0} = \langle\cos\left(n(\phi_{J,1}-\phi_{J,2}-\pi)\right)\rangle \equiv \langle\cos(n\varphi)\rangle \,,
\end{equation}
where $\phi_{J,1}$, $\phi_{J,2}$ are the azimuthal angles of the two jets.
The relative azimuthal angle $\varphi$ is defined such that $\varphi=0$ corresponds
to the back-to-back configuration.

The coefficients $\mathcal{C}_n$ can be expressed as 
\begin{eqnarray}
  \mathcal{C}_n &=& (4-3\delta_{n,0}) \intPS f(x_1) f(x_2) E_{n,\nu}(\veck_1) E^*_{n,\nu}(\veck_2) \non
   & & \times V(\veck_1,x_1) V(\veck_2,x_2) \cos(n\phi_{J2}) \cos(n\phi_{J1}) e^{\omega(n,\nu) Y}\,, \non
   \label{defCn}
\end{eqnarray}
where $Y=y_{J,1}-y_{J,2}$ and we have defined for brevity the integration over the phase space and over the parameter $\nu$ of conformal weight as
\begin{equation}
  \intPS=\int \dnu \dphij_1 \, \dk_1 \, \dx_1 \, \dphij_2 \, \dk_2 \, \dx_2\,,
\end{equation}
where $\nu$ is integrated from $-\infty$ to $+\infty$, $x_{1(2)}$ is integrated from $0$ to $1$ and $\phi_{J 1(2)}$ is integrated from $0$ to $2\pi$; $f$ are the usual parton distribution functions (PDFs) and $E_{n,\nu}$
are the LL BFKL eigenfunctions $E_{n,\nu}(\vecki) = \frac{1}
{\pi\sqrt{2}}\left(\vecki^2\right)^{i\nu-\frac{1}{2}}e^{in\phi_i}\,.$
The LL jet vertex reads
\begin{equation}
  V_{\rm a}^{(0)}(\veck,x) = \, 
  \frac{\alpha_s}{\sqrt{2}}\frac{C_{A/F}}{\veck^2}
  \delta\left(1-\frac{x_J}{x}\right)|\veckj|\delta^{(2)}(\veck-\veckj)\,,
\label{def-V0}
\end{equation}
where $C_A=N_c=3$ and $C_F=(N_c^2-1)/(2N_c)=4/3$ are to be used in the case of incoming gluon and quark respectively.
The jet vertex $V$ at NLL accuracy can be written as $V_{\rm a}(\veck,x) = 
V^{(0)}_{\rm a}(\veck,x) + \alpha_s V^{(1)}_{\rm a}(\veck,x)$.
The expression of $V^{(1)}_{\rm a}$, which has been recently reobtained in Ref.~\cite{Caporale:2011cc}, can be found in Ref.~\cite{Colferai:2010wu}.
Its expression in the limit of small cone jets has been computed in Ref.~\cite{Ivanov:2012ms} and used in Refs.~\cite{Caporale:2012ih,*Caporale:2013uva}.
It was also rederived within the high energy effective action
approach in Refs.~\cite{Hentschinski:2011tz,*Chachamis:2012cc}.

At NLL, the eigenvalue of the BFKL kernel is~\cite{Kotikov:2000pm,*Kotikov:2002ab,Ivanov:2005gn,Vera:2006un,*Vera:2007kn,*Schwennsen:2007hs}
\begin{equation}
  \omega(n,\nu) = \asbar \chi_0\left(|n|,\frac{1}{2}+i\nu\right) + \asbar^2
  \tilde{\chi}_1\left(|n|,\frac{1}{2}+i\nu\right)\,,
\end{equation}
where $\asbar = N_c\alpha_s/\pi$,
\begin{equation}
  \chi_0(n,\gamma) = 2\psi(1)-\psi\left(\gamma+\frac{n}
  {2}\right)-\psi\left(1-\gamma+\frac{n}{2}\right)\,,
\end{equation}
with $\psi(x) = \Gamma'(x) /\Gamma(x)$,
\begin{equation}
   \tilde{\chi}_1\left(n,\gamma\right) = \chi_1\left(n,\gamma\right)-
   \frac{\pi b_0}{N_c}\chi_0\left(n,\gamma\right)
   \ln\frac{|\veckjone|\cdot|\veckjtwo|}{\mu_R^2}\,,
\end{equation}
where the expression for $\chi_1$, which was obtained in
Refs.~\cite{Kotikov:2000pm,*Kotikov:2002ab}, can be found in Eq.~(2.17) of
Ref.~\cite{Ducloue:2013hia}.

\paragraph{BLM scale setting.}

The BLM procedure is a way of absorbing the non conformal
terms of the perturbative series in a redefinition of the coupling constant, to
improve the convergence of the perturbative series~\footnote{The BLM procedure was later extended to all orders, leading to the principle of maximal conformality (PMC) \cite{Brodsky:2011ig,*Brodsky:2011ta,*Brodsky:2012rj,*Brodsky:2012ik,*Mojaza:2012mf,*Wu:2013ei,*Brodsky:2013vpa,*Zheng:2013uja}.}.
In practice, one should extract the $\beta_0$-dependent part of an observable and choose the
renormalization scale to make it vanish.
The BLM procedure was first applied to BFKL dynamics in Refs.~\cite{Brodsky:1998kn,*Brodsky:2002ka} for the $\gamma^* \gamma^*$ total cross-section, considering the NLL corrections to the Green's function but using the LL $\gamma^*$ impact factor, with the important outcome of stabilizing the NLL BFKL intercept.
This method was used in a similar spirit in 
Refs.~\cite{Angioni:2011wj,*Hentschinski:2012kr,*Hentschinski:2013id}.
We follow the same line of thought, taking into account the NLL corrections to the jet vertex.

In the expression of the coefficients $\mathcal{C}_n$, the renormalization scale
$\mu_R$ enters both $\omega$ (through $\alpha_s$ and the second term of
$\tilde{\chi}_1$ which carries an explicit dependence on $\mu_R$) and $V$.
To separate the parts which depend on $\mu_R$ from those which do not, we rewrite
Eq.~(\ref{defCn}) as
\begin{eqnarray}
  \mathcal{C}_n &=& \alpha_s^2(4-3\delta_{n,0})\intPS D(\veck_1,x_1) D(\veck_2,x_2) \non
  & & \times A(x_1,\veck_1,\phi_{J1}) A^*(x_2,\veck_2,\phi_{J2}) e^{\omega(n,\nu) Y}\,,
  \label{Cnexpanded}
\end{eqnarray}
where $A(x_i,\veck_i,\phi_{Ji})=f(x_i)E_{n,\nu}(\veck_i)\cos(n\phi_{Ji})$. As
$V^{(0)}$ and $V^{(1)}$ both contain a global $\alpha_s$ factor, we have defined
$D^{(i)}(\veck,x)=V^{(i)}(\veck,x)/\alpha_s$ to make $D^{(0)}$ and $D^{(1)}$ 
$\alpha_s$-independent.
We now focus on the $\mu_R$-dependent part $D(\veck_1,x_1)
D(\veck_2,x_2)e^{\omega(n,\nu) Y} \equiv B_n$ of Eq.~(\ref{Cnexpanded}).
It can be expanded as the following series at NLL accuracy, for an arbitrary
renormalization scale $\mu_{R, \init}\,,$
\begin{widetext}
\begin{eqnarray}
  B_n &=& \left[ D^{(0)}(\veck_1,x_1)D^{(0)}(\veck_2,x_2) 
  +\alpha_s(\mu_{R, \init}) \left( D^{(1)}(\veck_1,x_1) D^{(0)}(\veck_2,x_2) 
  + D^{(0)}(\veck_1,x_1)D^{(1)}(\veck_2,x_2) \right) \right] \non
  & & \times \sum_{m=0}^\infty \frac{(\asbar(\mu_{R, \init}) \chi_0(n,\gamma) Y)^m}
  {m!} \left( 1+m \, \asbar(\mu_{R, \init}) \frac{\tilde{\chi}_1(n,\gamma}
  {\chi_0(n,\gamma} \right)\,.
  \label{amplitude}
\end{eqnarray}
\end{widetext}
Up to now, all the quantities we introduced were defined in the $\MSbar$ scheme.
However, the BLM procedure is more conveniently applied in a physical renormalization scheme, so we first perform the transition from the $\MSbar$ to the $\MOM$ scheme, which is equivalent to writing~\cite{Celmaster:1979dm,*Celmaster:1979km}
\begin{equation}
  \alpha_{\MSbar}=\alpha_{\MOM}\left(1+\alpha_{\MOM}\frac{T_{\MOM}}{\pi}\right)\,,
\end{equation}
where $T_{\MOM}=T_{\MOM}^\beta+T_{\MOM}^{conf}$,
\begin{eqnarray}
   T_{\MOM}^{conf} &=& \frac{N_c}{8}\left[\frac{17}{2}I+\frac{3}{2}\left(I-1\right)\xi+
   \left(1-\frac{1}{3}I\right)\xi^2-\frac{1}{6}\xi^3\right], \non
   T_{\MOM}^\beta &=& -\frac{\beta_0}{2} \left(1+\frac{2}{3}I\right),
\end{eqnarray}
where $\beta_0=(11N_c-2N_f)/3$, $N_f$ is the number of flavors, $I=-2\int_0^1 dx \ln(x)/[x^2-x+1] \simeq 2.3439$ and $\xi$ is a gauge parameter.
The variation of $B_n$ when going from the $\MSbar$ to the $\MOM$ scheme is
\begin{eqnarray}
  \delta B_n &=& D^{(0)}(\veck_1,x_1)D^{(0)}(\veck_2,x_2) \asbar(\mu_{R, \init}) \frac{T_{\MOM}}{N_c} \non
   & & \times \sum_{m=1}^\infty \frac{(\asbar(\mu_{R, \init}) \chi_0(n,\gamma) Y)^m}{(m-1)!}\,,
\end{eqnarray}
so that $B_{n,\MOM}=B_n+\delta B_n$. To express $B_{n,\MOM}$ as a function of an 
arbitrary renormalization scale $\mu_R$ we write $\alpha(\mu_{R, \init})$ as
\begin{equation}
  \alpha_s(\mu_{R, \init})=\alpha_s(\mu_R)\left(1-\alpha_s(\mu_R)\frac{\beta_0}
  {4\pi}\ln{\frac{{\mu^2_{R, \init}}}{\mu_R^2}}\right)\,.
  \label{runningalpha}
\end{equation}
We shall now insert Eq.~(\ref{runningalpha}) in the expression of $B_{n,\MOM}$ and
extract the $\beta_0$-dependent part.
One can see from the expression of  $V^{(1)}_{\rm a}$ given
in~\cite{Colferai:2010wu} that the term which depends on $\beta_0$ is proportional 
to the leading order part of the vertex, i.e. $D^{(1)\beta}(\veck_i,x_i)=
-\frac{\beta_0}{2\pi}\ln{\frac{\veck_i}{\mu_{R, \init}}}D^{(0)}(\veck_i,x_i)$.
Thus the part of $B_{n,\MOM}$ proportional to $\beta_0$ reads
\begin{widetext}
\begin{eqnarray}
  B_{n,\MOM}^\beta &=& D^{(0)}(\veck_1,x_1)D^{(0)}(\veck_2,x_2) \sum_{m=0}^\infty \alpha_s(\mu_R)^{m+1}
  \chi_0(n,\gamma)^m \left(\frac{Y N_c}{\pi}\right)^m \frac{1}{m!} \left[ -\frac{\beta_0}{2\pi}\ln{\frac{|\veck_1| \cdot |\veck_2|}{\mu^2_{R, \init}}} \right. \non
  & & + \left. m \frac{N_c}{\pi} \left( \frac{\tilde{\chi}_1^\beta(n,\gamma)}{\chi_0(n,\gamma)} + \frac{T_{\MOM}^\beta}{N_c} \right) - m\frac{\beta_0}{4\pi} \ln{\frac{\mu^2_{R, \init}}{\mu_R^2}} \right]\,,
\end{eqnarray}
\end{widetext}
where $\tilde{\chi}_1^\beta$ and $T_{\MOM}^\beta$ are the $\beta_0$-dependent
parts of $\tilde{\chi}_1$ and $T_{\MOM}$ respectively. The optimal scale
$\mu_{R,\BLM}$ is the value of $\mu_R$ that makes the expression inside the
brackets vanish. Taking into account the fact that our initial scale is
$\mu_{R, \init}=\sqrt{|\veckjone|\cdot|\veckjtwo|}$ and that
$D^{(0)}(\veck_i,x_i)$ contains a factor $\delta^{(2)}(\veck_i-\veckji)$ which
will enforce $|\veck_i|=|\veckji|$ after integrating over $\dk_i$, we need to
solve the equation
\begin{equation}
  \frac{N_c}{\pi} \left( \frac{\chi_1^\beta(n,\gamma)}{\chi_0(n,\gamma)} +
  \frac{T_{\MOM}^\beta}{N_c} \right) - \frac{\beta_0}{4\pi}
  \ln{\frac{|\veckjone|\cdot|\veckjtwo|}{\mu^2_{R,\BLM}}}=0\,,
\end{equation}
whose solution is 
\begin{equation}
  \mu^2_{R,\BLM}=|\veckjone|\cdot|\veckjtwo| \exp \left[ \frac{1}{2}
  \chi_0(n,\gamma)-\frac{5}{3}+2\left(\!1+\frac{2}{3}I\!\right)\! \right]\!.
\end{equation}

\paragraph{Theoretical uncertainties.}

Despite the fact that we have used the BLM procedure to fix the renormalization
scale, several theoretical uncertainties remain.

First,  the scale of the
prefactor $\alpha_s^2$ in Eq.~(\ref{Cnexpanded}) is not fixed in our
implementation of the scale fixing procedure.
To evaluate the corresponding uncertainty, we consider two cases, namely
either we take for this scale $\mu_{R,\BLM}$ or $\mu_{R,\init}\,.$

Second, our calculation involves the factorization scale $\mu_F$ which enters
both the PDFs and the hard part. In principle, one should vary independently
$\mu_R$ and $\mu_F$. But since the choice $\mu_R=\mu_F$ is made by all PDFs
fitting collaborations we are aware of, one could argue that, for consistency,
we should do the same. To estimate the reliability of our results, we did
two evaluations: one with $\mu_F=\mu_R=\mu_{R,\BLM}$ and one with the 'natural' choice
$\mu_F=\sqrt{|\veckjone|\cdot|\veckjtwo|}$.
In both cases, we  chose the single scale entering the  PDFs as $\mu_F$.

Third, several methods~\cite{Salam:1998tj,Ciafaloni:1998iv,*Ciafaloni:1999yw,*Ciafaloni:2003rd} 
have been proposed to improve the NLL BFKL Green's function by imposing its
compatibility with
DGLAP~\cite{Gribov:1972ri,*Lipatov:1974qm,*Altarelli:1977zs,*Dokshitzer:1977sg}
in the collinear limit. As in Ref.~\cite{Ducloue:2013hia}, we implemented scheme
3 of Ref.~\cite{Salam:1998tj} and found that the effect of such collinear
improvement was important for the cross section but much smaller than the two 
previous uncertainties for all the angular quantities we study here.

\paragraph{Results.}

Recently the CMS collaboration measured the azimuthal decorrelation of Mueller-Navelet
jets at the LHC at a center of mass energy of 7 TeV~\cite{CMS-PAS-FSQ-12-002}.
We here compare our results using the BLM procedure to this measurement. The 
quantities we discussed in the previous sections were differential with respect 
to the transverse momenta $\veckjone$, $\veckjtwo$ and the rapidities $y_{J,1}$, 
$y_{J,2}$ of the jets. Here we try to stay as close as possible to the 
configuration used in Ref.~\cite{CMS-PAS-FSQ-12-002}: $y_{J,1}$ and $y_{J,2}$ run 
between 0 and 4.7 and we integrate $\veckjone$ and $\veckjtwo$ from 35 GeV to 60 
GeV.
The CMS collaboration did not use an upper cut on the transverse 
momenta of the jets, but we have to do so for numerical reasons.
We have checked that our results do not depend strongly on the 
value of this cut, as the cross section is quickly decreasing with increasing transverse 
momenta.
We use the anti-$k_t$ jet algorithm~\cite{Cacciari:2008gp} with a size parameter
$R=0.5$ and the MSTW 2008 PDFs~\cite{Martin:2009iq}.
The results displayed in every figure include
the NLL BFKL calculation with the 'natural' choice $\mu_R=\mu_F=\sqrt{|\veckjone|\cdot|\veckjtwo|}$ (dashed line), 
the NLL BFKL calculation with the BLM scale choice (gray uncertainty band) and 
the CMS data (dots with error bars).
In our uncertainty band, we include the three effects discussed
in the previous section.

Before comparing our results with data, we would like to note that our calculation is performed at the partonic level and does not include hadronization effects. However the magnitude of these effects was estimated in~\cite{CMS-PAS-FSQ-12-002} to be smaller than the experimental uncertainties, which justifies this comparison. We also did not take into account multi-parton interactions, in which several partons from the same hadron take part in the interaction, as there is for now no theoretical framework to deal with such contributions at small $x$.

We first show results for the angular correlations $\avgcos$, $\avgcostwo$ and 
$\avgcosthree$ as a function of the relative rapidity $Y=y_{J,1}-y_{J,2}$ on 
Figs.~\ref{Fig:avgcos}, \ref{Fig:avgcos2} and \ref{Fig:avgcos3}, respectively. 
The conclusion for these three observables is similar: when one uses the 
'natural' scale, the NLL BFKL calculation is always above the data. But these 
data are much better described when setting the scale according to the BLM 
procedure.

\newcommand{\scal}{1}

\begin{figure}
  \includegraphics[scale=\scal]{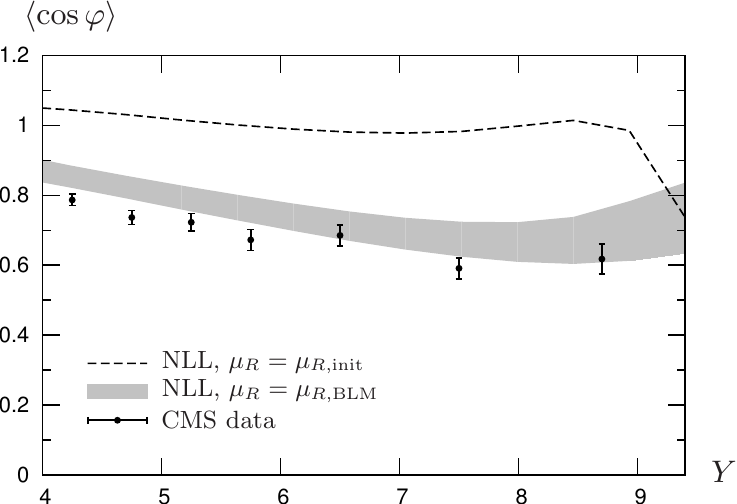}
  \caption{Variation of $\avgcos$ as a function of $Y$ at NLL accuracy compared with CMS data.}
  \label{Fig:avgcos}
\end{figure}
\begin{figure}
  \includegraphics[scale=\scal]{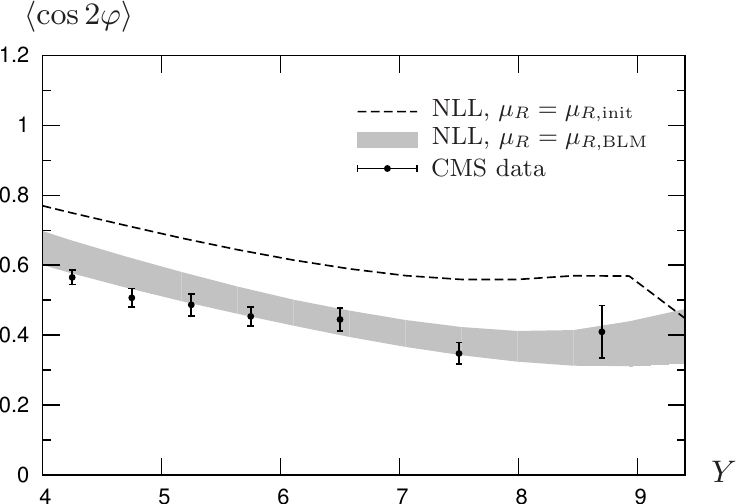}
  \caption{Variation of $\avgcostwo$ as a function of $Y$ at NLL accuracy compared with CMS data.}
  \label{Fig:avgcos2}
\end{figure}
\begin{figure}
  \includegraphics[scale=\scal]{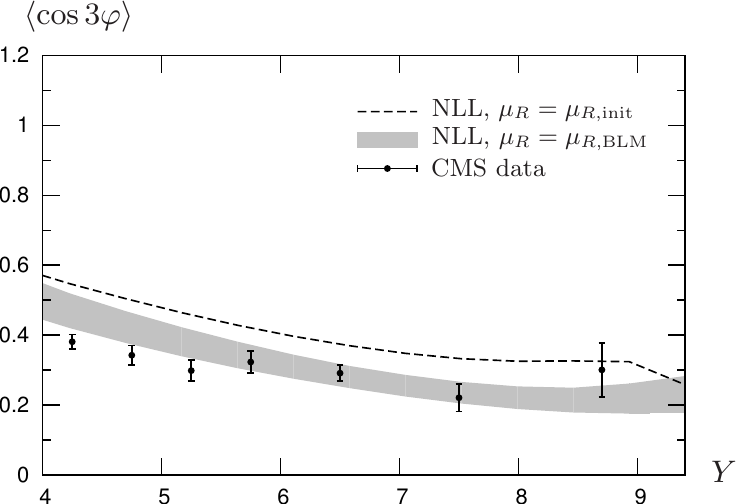}
  \caption{Variation of $\avgcosthree$ as a function of $Y$ at NLL accuracy compared with CMS data.}
  \label{Fig:avgcos3}
\end{figure}

\begin{figure}
  \includegraphics[scale=\scal]{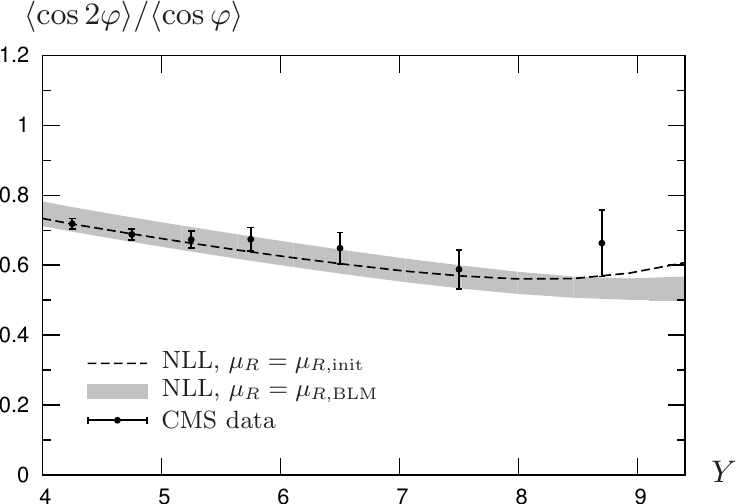}
  \caption{Variation of $\avgcostwo / \avgcos$ as a function of $Y$ at NLL
  accuracy compared with CMS data.}
  \label{Fig:avgcos2_avgcos}
\end{figure}

\begin{figure}
  \includegraphics[scale=\scal]{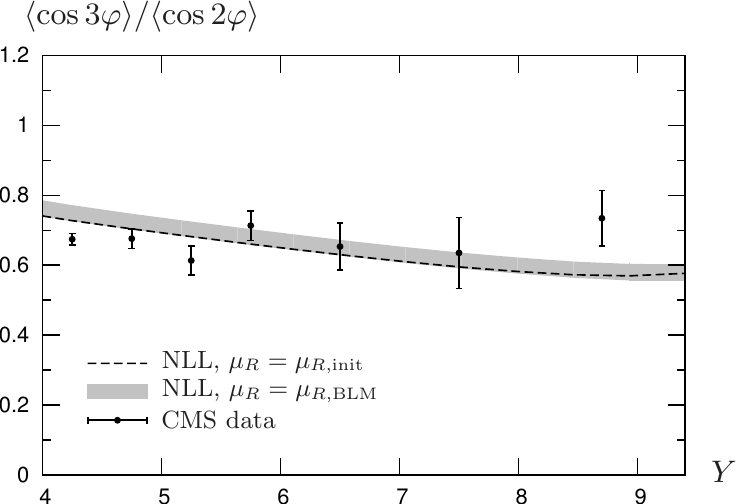}
  \caption{Variation of $\avgcosthree / \avgcostwo$ as a function of $Y$ at NLL
  accuracy compared with CMS data.}
  \label{Fig:avgcos3_avgcos2}
\end{figure}

On the other hand, the ratios $\avgcostwo/\avgcos$ and $\avgcosthree/\avgcostwo$
are almost not affected by the BLM procedure (see Figs.~\ref{Fig:avgcos2_avgcos}
and \ref{Fig:avgcos3_avgcos2}). This is because these observables are very stable
with respect to the scales, as was noticed before in
Refs.~\cite{Vera:2006un,*Vera:2007kn,*Schwennsen:2007hs,Colferai:2010wu,Ducloue:2013hia}.

\begin{figure}
  \includegraphics[scale=\scal]{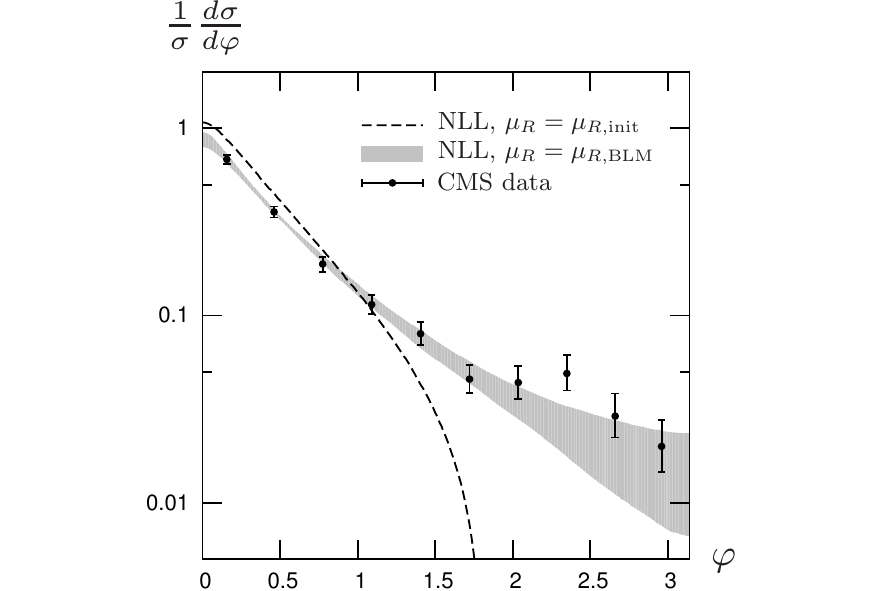}
  \caption{Azimuthal distribution at NLL accuracy compared with CMS data.}
  \label{Fig:deltaphi_dist}
\end{figure}

Another interesting observable measured in Ref.~\cite{CMS-PAS-FSQ-12-002} is the
azimuthal distribution of the jets $\frac{1}{{\sigma}}\frac{d{\sigma}}{d \varphi}$,
which can be expressed as
\begin{equation}
 \frac{1}{{\sigma}}\frac{d{\sigma}}{d \varphi}
  ~=~ \frac{1}{2\pi}
  \left\{1+2 \sum_{n=1}^\infty \cos{\left(n \varphi\right)}
  \left<\cos{\left( n \varphi \right)}\right>\right\}.
\end{equation}
In Fig.~\ref{Fig:deltaphi_dist} we show the comparison of our calculation with
the data for the azimuthal distribution integrated over the range $6.0<Y<9.4$.
We observe that using the 'natural' scale 
$\mu=\sqrt{|\veckjone|\cdot|\veckjtwo|}$, the BFKL calculation is slightly
above the data for $\varphi \lesssim 1$ and then becomes much lower than the
data, even reaching negative values for $\varphi \sim \pi$. This issue does
not arise when using BLM and the agreement with data then becomes very good
over the full $\varphi$ range.

\paragraph{Comparison with fixed-order.}

\begin{figure}
  \includegraphics[scale=\scal]{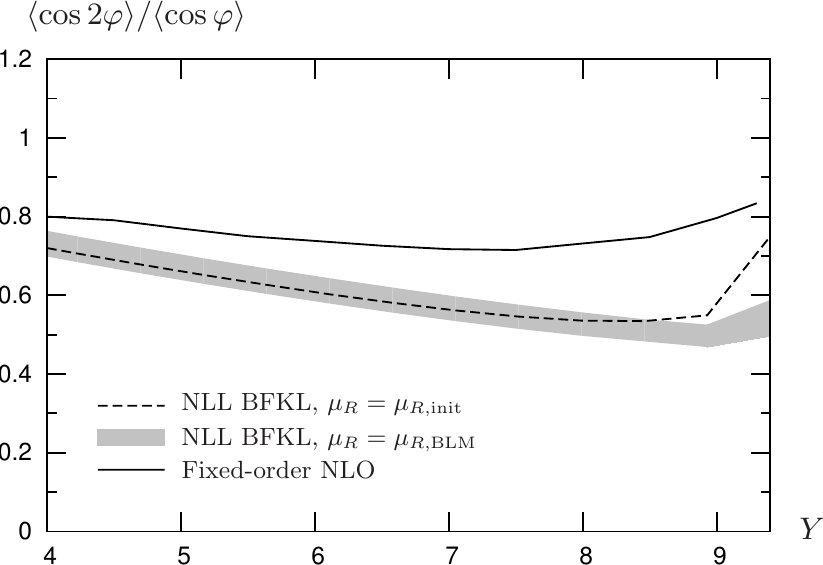}
  \caption{Variation of $\avgcostwo / \avgcos$ as a function of $Y$ at NLL
  accuracy compared with a fixed order treatment.}
  \label{Fig:avgcos2_avgcos_asym}
\end{figure}

Since the CMS collaboration considered configurations with identical lower cuts on the jets transverse momenta, which  would lead to unreliable results in a fixed-order treatment~\cite{Andersen:2001kta,*Fontannaz:2001nq}, a direct comparison of our analysis with this approach cannot be performed. In Fig.~\ref{Fig:avgcos2_avgcos_asym} we show the comparison of our BFKL calculation with the results obtained with the NLO fixed-order code Dijet~\cite{Aurenche:2008dn} for the ratio $\avgcostwo / \avgcos$ in the same kinematics as for previous results, but with the requirement that at least one jet has a transverse momentum larger than 50 GeV. As in~\cite{Ducloue:2013hia}, we see that there is a clear difference between BFKL and fixed-order so we expect that an experimental analysis in an asymmetric configuration would discriminate between these approaches.

\paragraph{Energy-momentum conservation.}

A general weakness of BFKL calculations is the absence of strict energy-momentum
conservation. This has been studied for Mueller-Navelet jets in the past~\cite{Orr:1997im,DelDuca:1994ng,*Marquet:2007xx}, using the leading order jet vertex.
These studies showed that this
is mainly an issue when $\veckjone$ and $\veckjtwo$ are different.
This effect should not be dramatic here, as we use the same lower cut on these variables
when comparing with CMS data and the cross section decreases quickly with increasing  $\veckjone$, $\veckjtwo$. Also we expect that the inclusion of the NLL corrections to the jet
vertex improves the situation.

\paragraph{Conclusions.}

In this work we have studied the azimuthal correlations of Mueller-Navelet jets
and compared the predictions of a full NLL BFKL calculation with data taken at
the LHC. We have shown that using the BLM procedure to fix the renormalization
scale leads to a very good agreement with the data, much better than when using
the 'natural' value $\sqrt{|\veckjone|\cdot|\veckjtwo|}$. 

\acknowledgments

We thank Fran\c cois Gelis and Edmond Iancu for warm hospitality at IPhT Saclay.
We thank Michel Fontannaz,  Grzegorz Brona, Hannes Jung, Victor Kim  and the Low-x
2013 Workshop participants for stimulating discussions.

This work is supported by the French Grant  PEPS-PTI,
the Polish Grant NCN No.~DEC-2011/01/B/ST2/03915 and  the Joint Research Activity 
Study of Strongly Interacting Matter (HadronPhysics3, Grant Agreement n.283286)
under the 7th Framework Programme of the European Community.


\begin{thebibliography}{10}%
\makeatletter
\providecommand \@ifxundefined [1]{%
 \ifx #1\undefined \expandafter \@firstoftwo
 \else \expandafter \@secondoftwo
\fi
}%
\providecommand \@ifnum [1]{%
 \ifnum #1\expandafter \@firstoftwo
 \else \expandafter \@secondoftwo
\fi
}%
\providecommand \enquote [1]{``#1''}%
\providecommand \bibnamefont  [1]{#1}%
\providecommand \bibfnamefont [1]{#1}%
\providecommand \citenamefont [1]{#1}%
\providecommand\href[0]{\@sanitize\@href}%
\providecommand\@href[1]{\endgroup\@@startlink{#1}\endgroup\@@href}%
\providecommand\@@href[1]{#1\@@endlink}%
\providecommand \@sanitize [0]{\begingroup\catcode`\&12\catcode`\#12\relax}%
\@ifxundefined \pdfoutput {\@firstoftwo}{%
 \@ifnum{\z@=\pdfoutput}{\@firstoftwo}{\@secondoftwo}%
}{%
 \providecommand\@@startlink[1]{\leavevmode}%
 \providecommand\@@endlink[0]{}%
}{%
 \providecommand\@@startlink[1]{%
  \leavevmode
  \pdfstartlink
   attr{/Border[0 0 1 ]/H/I/C[0 1 1]}%
   user{/Subtype/Link/A<</Type/Action/S/URI/URI(#1)>>}%
  \relax
 }%
 \providecommand\@@endlink[0]{\pdfendlink}%
}%
\providecommand \url  [0]{\begingroup\@sanitize \@url }%
\providecommand \@url [1]{\endgroup\@href {#1}{\urlprefix}}%
\providecommand \urlprefix [0]{URL }%
\providecommand \Eprint[0]{\href }%
\@ifxundefined \urlstyle {%
  \providecommand \doi [1]{doi:\discretionary{}{}{}#1}%
}{%
  \providecommand \doi [0]{doi:\discretionary{}{}{}\begingroup
  \urlstyle{rm}\Url }%
}%
\providecommand \doibase [0]{http://dx.doi.org/}%
\providecommand \Doi[1]{\href{\doibase#1}}%
\providecommand \bibAnnote [3]{%
  \BibitemShut{#1}%
  \begin{quotation}\noindent
    \textsc{Key:}\ #2\\\textsc{Annotation:}\ #3%
  \end{quotation}%
}%
\providecommand \bibAnnoteFile [2]{%
  \IfFileExists{#2}{\bibAnnote {#1} {#2} {\input{#2}}}{}%
}%
\providecommand \typeout [0]{\immediate \write \m@ne }%
\providecommand \selectlanguage [0]{\@gobble}%
\providecommand \bibinfo [0]{\@secondoftwo}%
\providecommand \bibfield [0]{\@secondoftwo}%
\providecommand \translation [1]{[#1]}%
\providecommand \BibitemOpen[0]{}%
\providecommand \bibitemStop [0]{}%
\providecommand \bibitemNoStop [0]{.\EOS\space}%
\providecommand \EOS [0]{\spacefactor3000\relax}%
\providecommand \BibitemShut [1]{\csname bibitem#1\endcsname}%
\bibitem{Fadin:1975cb}%
  \BibitemOpen
  \bibfield{author}{%
  \bibinfo {author} {\bibfnamefont{V.~S.}\ \bibnamefont{Fadin}}, \bibinfo
  {author} {\bibfnamefont{E.~A.}\ \bibnamefont{Kuraev}},\ and\ \bibinfo
  {author} {\bibfnamefont{L.~N.}\ \bibnamefont{Lipatov}},\ }%
  \bibfield{journal}{%
  \bibinfo {journal} {Phys. Lett.}\ }%
  \textbf{\bibinfo {volume} {B60}},\ \bibinfo {pages} {50} (\bibinfo {year}
  {1975})%
  \bibAnnoteFile{NoStop}{Fadin:1975cb}%
\bibitem{Kuraev:1976ge}%
  \BibitemOpen
  \bibfield{author}{%
  \bibinfo {author} {\bibfnamefont{E.~A.}\ \bibnamefont{Kuraev}}, \bibinfo
  {author} {\bibfnamefont{L.~N.}\ \bibnamefont{Lipatov}},\ and\ \bibinfo
  {author} {\bibfnamefont{V.~S.}\ \bibnamefont{Fadin}},\ }%
  \bibfield{journal}{%
  \bibinfo {journal} {Sov. Phys. JETP}\ }%
  \textbf{\bibinfo {volume} {44}},\ \bibinfo {pages} {443} (\bibinfo {year}
  {1976})%
  \bibAnnoteFile{NoStop}{Kuraev:1976ge}%
\bibitem{Kuraev:1977fs}%
  \BibitemOpen
  \bibfield{author}{%
  \bibinfo {author} {\bibfnamefont{E.~A.}\ \bibnamefont{Kuraev}}, \bibinfo
  {author} {\bibfnamefont{L.~N.}\ \bibnamefont{Lipatov}},\ and\ \bibinfo
  {author} {\bibfnamefont{V.~S.}\ \bibnamefont{Fadin}},\ }%
  \bibfield{journal}{%
  \bibinfo {journal} {Sov. Phys. JETP}\ }%
  \textbf{\bibinfo {volume} {45}},\ \bibinfo {pages} {199} (\bibinfo {year}
  {1977})%
  \bibAnnoteFile{NoStop}{Kuraev:1977fs}%
\bibitem{Balitsky:1978ic}%
  \BibitemOpen
  \bibfield{author}{%
  \bibinfo {author} {\bibfnamefont{I.~I.}\ \bibnamefont{Balitsky}}\ and\
  \bibinfo {author} {\bibfnamefont{L.~N.}\ \bibnamefont{Lipatov}},\ }%
  \bibfield{journal}{%
  \bibinfo {journal} {Sov. J. Nucl. Phys.}\ }%
  \textbf{\bibinfo {volume} {28}},\ \bibinfo {pages} {822} (\bibinfo {year}
  {1978})%
  \bibAnnoteFile{NoStop}{Balitsky:1978ic}%
\bibitem{Mueller:1986ey}%
  \BibitemOpen
  \bibfield{author}{%
  \bibinfo {author} {\bibfnamefont{A.~H.}\ \bibnamefont{Mueller}}\ and\
  \bibinfo {author} {\bibfnamefont{H.}~\bibnamefont{Navelet}},\ }%
  \bibfield{journal}{%
  \bibinfo {journal} {Nucl. Phys.}\ }%
  \textbf{\bibinfo {volume} {B282}},\ \bibinfo {pages} {727} (\bibinfo {year}
  {1987})%
  \bibAnnoteFile{NoStop}{Mueller:1986ey}%
\bibitem{CMS-PAS-FSQ-12-002}%
  \BibitemOpen
  \bibfield{author}{%
  \bibinfo {author} {\bibnamefont{{CMS Collaboration}}},\ }%
  \emph{\bibinfo {title} {{Azimuthal angle decorrelations of jets widely
  separated in rapidity in pp collisions at $\sqrt{s}=7$ TeV}}},\ \bibinfo
  {type} {CMS Physics Analysis Summary}\ \bibinfo {number}
  {CMS-PAS-FSQ-12-002}\ (\bibinfo {year} {2013})\
  \url{http://cds.cern.ch/record/1547075}%
  \bibAnnoteFile{NoStop}{CMS-PAS-FSQ-12-002}%
\bibitem{Fadin:1998py}%
  \BibitemOpen
  \bibfield{author}{%
  \bibinfo {author} {\bibfnamefont{V.~S.}\ \bibnamefont{Fadin}}\ and\ \bibinfo
  {author} {\bibfnamefont{L.~N.}\ \bibnamefont{Lipatov}},\ }%
  \bibfield{journal}{%
  \bibinfo {journal} {Phys. Lett.}\ }%
  \textbf{\bibinfo {volume} {B429}},\ \bibinfo {pages} {127} (\bibinfo {year}
  {1998}),\ \Eprint{http://arxiv.org/abs/hep-ph/9802290}{hep-ph/9802290}%
  \bibAnnoteFile{NoStop}{Fadin:1998py}%
\bibitem{Ciafaloni:1998gs}%
  \BibitemOpen
  \bibfield{author}{%
  \bibinfo {author} {\bibfnamefont{M.}~\bibnamefont{Ciafaloni}}\ and\ \bibinfo
  {author} {\bibfnamefont{G.}~\bibnamefont{Camici}},\ }%
  \bibfield{journal}{%
  \bibinfo {journal} {Phys. Lett.}\ }%
  \textbf{\bibinfo {volume} {B430}},\ \bibinfo {pages} {349} (\bibinfo {year}
  {1998}),\ \Eprint{http://arxiv.org/abs/hep-ph/9803389}{hep-ph/9803389}%
  \bibAnnoteFile{NoStop}{Ciafaloni:1998gs}%
\bibitem{Bartels:2001ge}%
  \BibitemOpen
  \bibfield{author}{%
  \bibinfo {author} {\bibfnamefont{J.}~\bibnamefont{Bartels}}, \bibinfo
  {author} {\bibfnamefont{D.}~\bibnamefont{Colferai}},\ and\ \bibinfo {author}
  {\bibfnamefont{G.~P.}\ \bibnamefont{Vacca}},\ }%
  \bibfield{journal}{%
  \bibinfo {journal} {Eur. Phys. J.}\ }%
  \textbf{\bibinfo {volume} {C24}},\ \bibinfo {pages} {83} (\bibinfo {year}
  {2002}),\ \Eprint{http://arxiv.org/abs/hep-ph/0112283}{hep-ph/0112283}%
  \bibAnnoteFile{NoStop}{Bartels:2001ge}%
\bibitem{Bartels:2002yj}%
  \BibitemOpen
  \bibfield{author}{%
  \bibinfo {author} {\bibfnamefont{J.}~\bibnamefont{Bartels}}, \bibinfo
  {author} {\bibfnamefont{D.}~\bibnamefont{Colferai}},\ and\ \bibinfo {author}
  {\bibfnamefont{G.~P.}\ \bibnamefont{Vacca}},\ }%
  \bibfield{journal}{%
  \bibinfo {journal} {Eur. Phys. J.}\ }%
  \textbf{\bibinfo {volume} {C29}},\ \bibinfo {pages} {235} (\bibinfo {year}
  {2003}),\ \Eprint{http://arxiv.org/abs/hep-ph/0206290}{hep-ph/0206290}%
  \bibAnnoteFile{NoStop}{Bartels:2002yj}%
\bibitem{Colferai:2010wu}%
  \BibitemOpen
  \bibfield{author}{%
  \bibinfo {author} {\bibfnamefont{D.}~\bibnamefont{Colferai}}, \bibinfo
  {author} {\bibfnamefont{F.}~\bibnamefont{Schwennsen}}, \bibinfo {author}
  {\bibfnamefont{L.}~\bibnamefont{Szymanowski}},\ and\ \bibinfo {author}
  {\bibfnamefont{S.}~\bibnamefont{Wallon}},\ }%
  \bibfield{journal}{%
  \Doi{10.1007/JHEP12(2010)026}{\bibinfo {journal} {JHEP}}\ }%
  \textbf{\bibinfo {volume} {12}},\ \bibinfo {pages} {026} (\bibinfo {year}
  {2010}),\ \Eprint{http://arxiv.org/abs/1002.1365}{arXiv:1002.1365 [hep-ph]}%
  \bibAnnoteFile{NoStop}{Colferai:2010wu}%
\bibitem{Ducloue:2013hia}%
  \BibitemOpen
  \bibfield{author}{%
  \bibinfo {author} {\bibfnamefont{B.}~\bibnamefont{Duclou\'e}}, \bibinfo
  {author} {\bibfnamefont{L.}~\bibnamefont{Szymanowski}},\ and\ \bibinfo
  {author} {\bibfnamefont{S.}~\bibnamefont{Wallon}},\ }%
  \bibfield{journal}{%
  \Doi{10.1007/JHEP05(2013)096}{\bibinfo {journal} {JHEP}}\ }%
  \textbf{\bibinfo {volume} {1305}},\ \bibinfo {pages} {096} (\bibinfo {year}
  {2013}),\ \Eprint{http://arxiv.org/abs/1302.7012}{arXiv:1302.7012 [hep-ph]}%
  \bibAnnoteFile{NoStop}{Ducloue:2013hia}%
\bibitem{Brodsky:1982gc}%
  \BibitemOpen
  \bibfield{author}{%
  \bibinfo {author} {\bibfnamefont{S.~J.}\ \bibnamefont{Brodsky}}, \bibinfo
  {author} {\bibfnamefont{G.~P.}\ \bibnamefont{Lepage}},\ and\ \bibinfo
  {author} {\bibfnamefont{P.~B.}\ \bibnamefont{Mackenzie}},\ }%
  \bibfield{journal}{%
  \bibinfo {journal} {Phys. Rev.}\ }%
  \textbf{\bibinfo {volume} {D28}},\ \bibinfo {pages} {228} (\bibinfo {year}
  {1983})%
  \bibAnnoteFile{NoStop}{Brodsky:1982gc}%
\bibitem{Brodsky:1998kn}%
  \BibitemOpen
  \bibfield{author}{%
  \bibinfo {author} {\bibfnamefont{S.~J.}\ \bibnamefont{Brodsky}}, \bibinfo
  {author} {\bibfnamefont{V.~S.}\ \bibnamefont{Fadin}}, \bibinfo {author}
  {\bibfnamefont{V.~T.}\ \bibnamefont{Kim}}, \bibinfo {author}
  {\bibfnamefont{L.~N.}\ \bibnamefont{Lipatov}},\ and\ \bibinfo {author}
  {\bibfnamefont{G.~B.}\ \bibnamefont{Pivovarov}},\ }%
  \bibfield{journal}{%
  \bibinfo {journal} {JETP Lett.}\ }%
  \textbf{\bibinfo {volume} {70}},\ \bibinfo {pages} {155} (\bibinfo {year}
  {1999}),\ \Eprint{http://arxiv.org/abs/hep-ph/9901229}{hep-ph/9901229}%
  \bibAnnoteFile{NoStop}{Brodsky:1998kn}%
\bibitem{Brodsky:2002ka}%
  \BibitemOpen
  \bibfield{author}{%
  \bibinfo {author} {\bibfnamefont{S.~J.}\ \bibnamefont{Brodsky}}, \bibinfo
  {author} {\bibfnamefont{V.~S.}\ \bibnamefont{Fadin}}, \bibinfo {author}
  {\bibfnamefont{V.~T.}\ \bibnamefont{Kim}}, \bibinfo {author}
  {\bibfnamefont{L.~N.}\ \bibnamefont{Lipatov}},\ and\ \bibinfo {author}
  {\bibfnamefont{G.~B.}\ \bibnamefont{Pivovarov}},\ }%
  \bibfield{journal}{%
  \bibinfo {journal} {JETP Lett.}\ }%
  \textbf{\bibinfo {volume} {76}},\ \bibinfo {pages} {249} (\bibinfo {year}
  {2002}),\ \Eprint{http://arxiv.org/abs/hep-ph/0207297}{hep-ph/0207297}%
  \bibAnnoteFile{NoStop}{Brodsky:2002ka}%
\bibitem{DelDuca:1993mn}%
  \BibitemOpen
  \bibfield{author}{%
  \bibinfo {author} {\bibfnamefont{V.}~\bibnamefont{Del~Duca}}\ and\ \bibinfo
  {author} {\bibfnamefont{C.~R.}\ \bibnamefont{Schmidt}},\ }%
  \bibfield{journal}{%
  \Doi{10.1103/PhysRevD.49.4510}{\bibinfo {journal} {Phys. Rev.}}\ }%
  \textbf{\bibinfo {volume} {D49}},\ \bibinfo {pages} {4510} (\bibinfo {year}
  {1994}),\ \Eprint{http://arxiv.org/abs/hep-ph/9311290}{hep-ph/9311290}%
  \bibAnnoteFile{NoStop}{DelDuca:1993mn}%
\bibitem{Stirling:1994zs}%
  \BibitemOpen
  \bibfield{author}{%
  \bibinfo {author} {\bibfnamefont{W.~J.}\ \bibnamefont{Stirling}},\ }%
  \bibfield{journal}{%
  \bibinfo {journal} {Nucl. Phys.}\ }%
  \textbf{\bibinfo {volume} {B423}},\ \bibinfo {pages} {56} (\bibinfo {year}
  {1994}),\ \Eprint{http://arxiv.org/abs/hep-ph/9401266}{hep-ph/9401266}%
  \bibAnnoteFile{NoStop}{Stirling:1994zs}%
\bibitem{Caporale:2011cc}%
  \BibitemOpen
  \bibfield{author}{%
  \bibinfo {author} {\bibfnamefont{F.}~\bibnamefont{Caporale}}, \bibinfo
  {author} {\bibfnamefont{D.~Y.}\ \bibnamefont{Ivanov}}, \bibinfo {author}
  {\bibfnamefont{B.}~\bibnamefont{Murdaca}}, \bibinfo {author}
  {\bibfnamefont{A.}~\bibnamefont{Papa}},\ and\ \bibinfo {author}
  {\bibfnamefont{A.}~\bibnamefont{Perri}},\ }%
  \bibfield{journal}{%
  \Doi{10.1007/JHEP02(2012)101}{\bibinfo {journal} {JHEP}}\ }%
  \textbf{\bibinfo {volume} {1202}},\ \bibinfo {pages} {101} (\bibinfo {year}
  {2012}),\ \Eprint{http://arxiv.org/abs/1112.3752}{arXiv:1112.3752 [hep-ph]}%
  \bibAnnoteFile{NoStop}{Caporale:2011cc}%
\bibitem{Ivanov:2012ms}%
  \BibitemOpen
  \bibfield{author}{%
  \bibinfo {author} {\bibfnamefont{D.~Y.}\ \bibnamefont{Ivanov}}\ and\ \bibinfo
  {author} {\bibfnamefont{A.}~\bibnamefont{Papa}},\ }%
  \bibfield{journal}{%
  \Doi{10.1007/JHEP05(2012)086}{\bibinfo {journal} {JHEP}}\ }%
  \textbf{\bibinfo {volume} {1205}},\ \bibinfo {pages} {086} (\bibinfo {year}
  {2012}),\ \Eprint{http://arxiv.org/abs/1202.1082}{arXiv:1202.1082 [hep-ph]}%
  \bibAnnoteFile{NoStop}{Ivanov:2012ms}%
\bibitem{Caporale:2012ih}%
  \BibitemOpen
  \bibfield{author}{%
  \bibinfo {author} {\bibfnamefont{F.}~\bibnamefont{Caporale}}, \bibinfo
  {author} {\bibfnamefont{D.~Y.}\ \bibnamefont{Ivanov}}, \bibinfo {author}
  {\bibfnamefont{B.}~\bibnamefont{Murdaca}},\ and\ \bibinfo {author}
  {\bibfnamefont{A.}~\bibnamefont{Papa}}}%
   (\bibinfo {year} {2012}),\
  \Eprint{http://arxiv.org/abs/1211.7225}{arXiv:1211.7225 [hep-ph]}%
  \bibAnnoteFile{NoStop}{Caporale:2012ih}%
\bibitem{Caporale:2013uva}%
  \BibitemOpen
  \bibfield{author}{%
  \bibinfo {author} {\bibfnamefont{F.}~\bibnamefont{Caporale}}, \bibinfo
  {author} {\bibfnamefont{B.}~\bibnamefont{Murdaca}}, \bibinfo {author}
  {\bibfnamefont{A.}~\bibnamefont{Sabio~Vera}},\ and\ \bibinfo {author}
  {\bibfnamefont{C.}~\bibnamefont{Salas}},\ }%
  \bibfield{journal}{%
  \Doi{10.1016/j.nuclphysb.2013.07.005}{\bibinfo {journal} {Nucl. Phys.}}\ }%
  \textbf{\bibinfo {volume} {B875}},\ \bibinfo {pages} {134} (\bibinfo {year}
  {2013}),\ \Eprint{http://arxiv.org/abs/1305.4620}{arXiv:1305.4620 [hep-ph]}%
  \bibAnnoteFile{NoStop}{Caporale:2013uva}%
\bibitem{Hentschinski:2011tz}%
  \BibitemOpen
  \bibfield{author}{%
  \bibinfo {author} {\bibfnamefont{M.}~\bibnamefont{Hentschinski}}\ and\
  \bibinfo {author} {\bibfnamefont{A.}~\bibnamefont{Sabio~Vera}},\ }%
  \bibfield{journal}{%
  \Doi{10.1103/PhysRevD.85.056006}{\bibinfo {journal} {Phys. Rev.}}\ }%
  \textbf{\bibinfo {volume} {D85}},\ \bibinfo {pages} {056006} (\bibinfo {year}
  {2012}),\ \Eprint{http://arxiv.org/abs/1110.6741}{arXiv:1110.6741 [hep-ph]}%
  \bibAnnoteFile{NoStop}{Hentschinski:2011tz}%
\bibitem{Chachamis:2012cc}%
  \BibitemOpen
  \bibfield{author}{%
  \bibinfo {author} {\bibfnamefont{G.}~\bibnamefont{Chachamis}}, \bibinfo
  {author} {\bibfnamefont{M.}~\bibnamefont{Hentschinski}}, \bibinfo {author}
  {\bibfnamefont{J.~D.}\ \bibnamefont{Madrigal}},\ and\ \bibinfo {author}
  {\bibfnamefont{A.}~\bibnamefont{Sabio~Vera}},\ }%
  \bibfield{journal}{%
  \Doi{10.1103/PhysRevD.87.076009}{\bibinfo {journal} {Phys. Rev.}}\ }%
  \textbf{\bibinfo {volume} {D87}},\ \bibinfo {pages} {076009} (\bibinfo {year}
  {2013}),\ \Eprint{http://arxiv.org/abs/1212.4992}{arXiv:1212.4992 [hep-ph]}%
  \bibAnnoteFile{NoStop}{Chachamis:2012cc}%
\bibitem{Kotikov:2000pm}%
  \BibitemOpen
  \bibfield{author}{%
  \bibinfo {author} {\bibfnamefont{A.~V.}\ \bibnamefont{Kotikov}}\ and\
  \bibinfo {author} {\bibfnamefont{L.~N.}\ \bibnamefont{Lipatov}},\ }%
  \bibfield{journal}{%
  \bibinfo {journal} {Nucl. Phys.}\ }%
  \textbf{\bibinfo {volume} {B582}},\ \bibinfo {pages} {19} (\bibinfo {year}
  {2000}),\ \Eprint{http://arxiv.org/abs/hep-ph/0004008}{hep-ph/0004008}%
  \bibAnnoteFile{NoStop}{Kotikov:2000pm}%
\bibitem{Kotikov:2002ab}%
  \BibitemOpen
  \bibfield{author}{%
  \bibinfo {author} {\bibfnamefont{A.~V.}\ \bibnamefont{Kotikov}}\ and\
  \bibinfo {author} {\bibfnamefont{L.~N.}\ \bibnamefont{Lipatov}},\ }%
  \bibfield{journal}{%
  \Doi{10.1016/S0550-3213(03)00264-5}{\bibinfo {journal} {Nucl. Phys.}}\ }%
  \textbf{\bibinfo {volume} {B661}},\ \bibinfo {pages} {19} (\bibinfo {year}
  {2003}),\ \Eprint{http://arxiv.org/abs/hep-ph/0208220}{hep-ph/0208220}%
  \bibAnnoteFile{NoStop}{Kotikov:2002ab}%
\bibitem{Ivanov:2005gn}%
  \BibitemOpen
  \bibfield{author}{%
  \bibinfo {author} {\bibfnamefont{D.~Y.}\ \bibnamefont{Ivanov}}\ and\ \bibinfo
  {author} {\bibfnamefont{A.}~\bibnamefont{Papa}},\ }%
  \bibfield{journal}{%
  \bibinfo {journal} {Nucl. Phys.}\ }%
  \textbf{\bibinfo {volume} {B732}},\ \bibinfo {pages} {183} (\bibinfo {year}
  {2006}),\ \Eprint{http://arxiv.org/abs/hep-ph/0508162}{hep-ph/0508162}%
  \bibAnnoteFile{NoStop}{Ivanov:2005gn}%
\bibitem{Vera:2006un}%
  \BibitemOpen
  \bibfield{author}{%
  \bibinfo {author} {\bibfnamefont{A.}~\bibnamefont{Sabio~Vera}},\ }%
  \bibfield{journal}{%
  \bibinfo {journal} {Nucl. Phys.}\ }%
  \textbf{\bibinfo {volume} {B746}},\ \bibinfo {pages} {1} (\bibinfo {year}
  {2006}),\ \Eprint{http://arxiv.org/abs/hep-ph/0602250}{hep-ph/0602250}%
  \bibAnnoteFile{NoStop}{Vera:2006un}%
\bibitem{Vera:2007kn}%
  \BibitemOpen
  \bibfield{author}{%
  \bibinfo {author} {\bibfnamefont{A.}~\bibnamefont{Sabio~Vera}}\ and\ \bibinfo
  {author} {\bibfnamefont{F.}~\bibnamefont{Schwennsen}},\ }%
  \bibfield{journal}{%
  \bibinfo {journal} {Nucl. Phys.}\ }%
  \textbf{\bibinfo {volume} {B776}},\ \bibinfo {pages} {170} (\bibinfo {year}
  {2007}),\ \Eprint{http://arxiv.org/abs/hep-ph/0702158}{hep-ph/0702158}%
  \bibAnnoteFile{NoStop}{Vera:2007kn}%
\bibitem{Schwennsen:2007hs}%
  \BibitemOpen
  \bibfield{author}{%
  \bibinfo {author} {\bibfnamefont{F.}~\bibnamefont{Schwennsen}}}%
   (\bibinfo {year} {2007}),\
  \Eprint{http://arxiv.org/abs/hep-ph/0703198}{hep-ph/0703198}%
  \bibAnnoteFile{NoStop}{Schwennsen:2007hs}%
\bibitem{Note1}%
  \BibitemOpen
  \bibinfo {note} {The BLM procedure was later extended to all orders, leading
  to the principle of maximal conformality (PMC) \cite
  {Brodsky:2011ig,*Brodsky:2011ta,*Brodsky:2012rj,*Brodsky:2012ik,*Mojaza:2012%
mf,*Wu:2013ei,*Brodsky:2013vpa,*Zheng:2013uja}.}%
  \bibAnnoteFile{Stop}{Note1}%
\bibitem{Angioni:2011wj}%
  \BibitemOpen
  \bibfield{author}{%
  \bibinfo {author} {\bibfnamefont{M.}~\bibnamefont{Angioni}}, \bibinfo
  {author} {\bibfnamefont{G.}~\bibnamefont{Chachamis}}, \bibinfo {author}
  {\bibfnamefont{J.}~\bibnamefont{Madrigal}},\ and\ \bibinfo {author}
  {\bibfnamefont{A.}~\bibnamefont{Sabio~Vera}},\ }%
  \bibfield{journal}{%
  \Doi{10.1103/PhysRevLett.107.191601}{\bibinfo {journal} {Phys. Rev. Lett.}}\
  }%
  \textbf{\bibinfo {volume} {107}},\ \bibinfo {pages} {191601} (\bibinfo {year}
  {2011}),\ \Eprint{http://arxiv.org/abs/1106.6172}{arXiv:1106.6172 [hep-th]}%
  \bibAnnoteFile{NoStop}{Angioni:2011wj}%
\bibitem{Hentschinski:2012kr}%
  \BibitemOpen
  \bibfield{author}{%
  \bibinfo {author} {\bibfnamefont{M.}~\bibnamefont{Hentschinski}}, \bibinfo
  {author} {\bibfnamefont{A.}~\bibnamefont{Sabio~Vera}},\ and\ \bibinfo
  {author} {\bibfnamefont{C.}~\bibnamefont{Salas}},\ }%
  \bibfield{journal}{%
  \Doi{10.1103/PhysRevLett.110.041601}{\bibinfo {journal} {Phys. Rev. Lett.}}\
  }%
  \textbf{\bibinfo {volume} {110}},\ \bibinfo {pages} {041601} (\bibinfo {year}
  {2013}),\ \Eprint{http://arxiv.org/abs/1209.1353}{arXiv:1209.1353 [hep-ph]}%
  \bibAnnoteFile{NoStop}{Hentschinski:2012kr}%
\bibitem{Hentschinski:2013id}%
  \BibitemOpen
  \bibfield{author}{%
  \bibinfo {author} {\bibfnamefont{M.}~\bibnamefont{Hentschinski}}, \bibinfo
  {author} {\bibfnamefont{A.}~\bibnamefont{Sabio~Vera}},\ and\ \bibinfo
  {author} {\bibfnamefont{C.}~\bibnamefont{Salas}},\ }%
  \bibfield{journal}{%
  \Doi{10.1103/PhysRevD.87.076005}{\bibinfo {journal} {Phys. Rev.}}\ }%
  \textbf{\bibinfo {volume} {D87}},\ \bibinfo {pages} {076005} (\bibinfo {year}
  {2013}),\ \Eprint{http://arxiv.org/abs/1301.5283}{arXiv:1301.5283 [hep-ph]}%
  \bibAnnoteFile{NoStop}{Hentschinski:2013id}%
\bibitem{Celmaster:1979dm}%
  \BibitemOpen
  \bibfield{author}{%
  \bibinfo {author} {\bibfnamefont{W.}~\bibnamefont{Celmaster}}\ and\ \bibinfo
  {author} {\bibfnamefont{R.~J.}\ \bibnamefont{Gonsalves}},\ }%
  \bibfield{journal}{%
  \Doi{10.1103/PhysRevLett.42.1435}{\bibinfo {journal} {Phys. Rev. Lett.}}\ }%
  \textbf{\bibinfo {volume} {42}},\ \bibinfo {pages} {1435} (\bibinfo {year}
  {1979})%
  \bibAnnoteFile{NoStop}{Celmaster:1979dm}%
\bibitem{Celmaster:1979km}%
  \BibitemOpen
  \bibfield{author}{%
  \bibinfo {author} {\bibfnamefont{W.}~\bibnamefont{Celmaster}}\ and\ \bibinfo
  {author} {\bibfnamefont{R.~J.}\ \bibnamefont{Gonsalves}},\ }%
  \bibfield{journal}{%
  \Doi{10.1103/PhysRevD.20.1420}{\bibinfo {journal} {Phys. Rev.}}\ }%
  \textbf{\bibinfo {volume} {D20}},\ \bibinfo {pages} {1420} (\bibinfo {year}
  {1979})%
  \bibAnnoteFile{NoStop}{Celmaster:1979km}%
\bibitem{Salam:1998tj}%
  \BibitemOpen
  \bibfield{author}{%
  \bibinfo {author} {\bibfnamefont{G.~P.}\ \bibnamefont{Salam}},\ }%
  \bibfield{journal}{%
  \bibinfo {journal} {JHEP}\ }%
  \textbf{\bibinfo {volume} {07}},\ \bibinfo {pages} {019} (\bibinfo {year}
  {1998}),\ \Eprint{http://arxiv.org/abs/hep-ph/9806482}{hep-ph/9806482}%
  \bibAnnoteFile{NoStop}{Salam:1998tj}%
\bibitem{Ciafaloni:1998iv}%
  \BibitemOpen
  \bibfield{author}{%
  \bibinfo {author} {\bibfnamefont{M.}~\bibnamefont{Ciafaloni}}\ and\ \bibinfo
  {author} {\bibfnamefont{D.}~\bibnamefont{Colferai}},\ }%
  \bibfield{journal}{%
  \bibinfo {journal} {Phys. Lett.}\ }%
  \textbf{\bibinfo {volume} {B452}},\ \bibinfo {pages} {372} (\bibinfo {year}
  {1999}),\ \Eprint{http://arxiv.org/abs/hep-ph/9812366}{hep-ph/9812366}%
  \bibAnnoteFile{NoStop}{Ciafaloni:1998iv}%
\bibitem{Ciafaloni:1999yw}%
  \BibitemOpen
  \bibfield{author}{%
  \bibinfo {author} {\bibfnamefont{M.}~\bibnamefont{Ciafaloni}}, \bibinfo
  {author} {\bibfnamefont{D.}~\bibnamefont{Colferai}},\ and\ \bibinfo {author}
  {\bibfnamefont{G.~P.}\ \bibnamefont{Salam}},\ }%
  \bibfield{journal}{%
  \bibinfo {journal} {Phys. Rev.}\ }%
  \textbf{\bibinfo {volume} {D60}},\ \bibinfo {pages} {114036} (\bibinfo {year}
  {1999}),\ \Eprint{http://arxiv.org/abs/hep-ph/9905566}{hep-ph/9905566}%
  \bibAnnoteFile{NoStop}{Ciafaloni:1999yw}%
\bibitem{Ciafaloni:2003rd}%
  \BibitemOpen
  \bibfield{author}{%
  \bibinfo {author} {\bibfnamefont{M.}~\bibnamefont{Ciafaloni}}, \bibinfo
  {author} {\bibfnamefont{D.}~\bibnamefont{Colferai}}, \bibinfo {author}
  {\bibfnamefont{G.~P.}\ \bibnamefont{Salam}},\ and\ \bibinfo {author}
  {\bibfnamefont{A.~M.}\ \bibnamefont{Stasto}},\ }%
  \bibfield{journal}{%
  \bibinfo {journal} {Phys. Rev.}\ }%
  \textbf{\bibinfo {volume} {D68}},\ \bibinfo {pages} {114003} (\bibinfo {year}
  {2003}),\ \Eprint{http://arxiv.org/abs/hep-ph/0307188}{hep-ph/0307188}%
  \bibAnnoteFile{NoStop}{Ciafaloni:2003rd}%
\bibitem{Gribov:1972ri}%
  \BibitemOpen
  \bibfield{author}{%
  \bibinfo {author} {\bibfnamefont{V.~N.}\ \bibnamefont{Gribov}}\ and\ \bibinfo
  {author} {\bibfnamefont{L.~N.}\ \bibnamefont{Lipatov}},\ }%
  \bibfield{journal}{%
  \bibinfo {journal} {Sov. J. Nucl. Phys.}\ }%
  \textbf{\bibinfo {volume} {15}},\ \bibinfo {pages} {438} (\bibinfo {year}
  {1972})%
  \bibAnnoteFile{NoStop}{Gribov:1972ri}%
\bibitem{Lipatov:1974qm}%
  \BibitemOpen
  \bibfield{author}{%
  \bibinfo {author} {\bibfnamefont{L.~N.}\ \bibnamefont{Lipatov}},\ }%
  \bibfield{journal}{%
  \bibinfo {journal} {Sov. J. Nucl. Phys.}\ }%
  \textbf{\bibinfo {volume} {20}},\ \bibinfo {pages} {94} (\bibinfo {year}
  {1975})%
  \bibAnnoteFile{NoStop}{Lipatov:1974qm}%
\bibitem{Altarelli:1977zs}%
  \BibitemOpen
  \bibfield{author}{%
  \bibinfo {author} {\bibfnamefont{G.}~\bibnamefont{Altarelli}}\ and\ \bibinfo
  {author} {\bibfnamefont{G.}~\bibnamefont{Parisi}},\ }%
  \bibfield{journal}{%
  \bibinfo {journal} {Nucl. Phys.}\ }%
  \textbf{\bibinfo {volume} {B126}},\ \bibinfo {pages} {298} (\bibinfo {year}
  {1977})%
  \bibAnnoteFile{NoStop}{Altarelli:1977zs}%
\bibitem{Dokshitzer:1977sg}%
  \BibitemOpen
  \bibfield{author}{%
  \bibinfo {author} {\bibfnamefont{Y.~L.}\ \bibnamefont{Dokshitzer}},\ }%
  \bibfield{journal}{%
  \bibinfo {journal} {Sov. Phys. JETP}\ }%
  \textbf{\bibinfo {volume} {46}},\ \bibinfo {pages} {641} (\bibinfo {year}
  {1977})%
  \bibAnnoteFile{NoStop}{Dokshitzer:1977sg}%
\bibitem{Cacciari:2008gp}%
  \BibitemOpen
  \bibfield{author}{%
  \bibinfo {author} {\bibfnamefont{M.}~\bibnamefont{Cacciari}}, \bibinfo
  {author} {\bibfnamefont{G.~P.}\ \bibnamefont{Salam}},\ and\ \bibinfo {author}
  {\bibfnamefont{G.}~\bibnamefont{Soyez}},\ }%
  \bibfield{journal}{%
  \Doi{10.1088/1126-6708/2008/04/063}{\bibinfo {journal} {JHEP}}\ }%
  \textbf{\bibinfo {volume} {0804}},\ \bibinfo {pages} {063} (\bibinfo {year}
  {2008}),\ \Eprint{http://arxiv.org/abs/0802.1189}{arXiv:0802.1189 [hep-ph]}%
  \bibAnnoteFile{NoStop}{Cacciari:2008gp}%
\bibitem{Martin:2009iq}%
  \BibitemOpen
  \bibfield{author}{%
  \bibinfo {author} {\bibfnamefont{A.~D.}\ \bibnamefont{Martin}}, \bibinfo
  {author} {\bibfnamefont{W.~J.}\ \bibnamefont{Stirling}}, \bibinfo {author}
  {\bibfnamefont{R.~S.}\ \bibnamefont{Thorne}},\ and\ \bibinfo {author}
  {\bibfnamefont{G.}~\bibnamefont{Watt}},\ }%
  \bibfield{journal}{%
  \Doi{10.1140/epjc/s10052-009-1072-5}{\bibinfo {journal} {Eur. Phys. J.}}\ }%
  \textbf{\bibinfo {volume} {C63}},\ \bibinfo {pages} {189} (\bibinfo {year}
  {2009}),\ \Eprint{http://arxiv.org/abs/0901.0002}{arXiv:0901.0002 [hep-ph]}%
  \bibAnnoteFile{NoStop}{Martin:2009iq}%
\bibitem{Andersen:2001kta}%
  \BibitemOpen
  \bibfield{author}{%
  \bibinfo {author} {\bibfnamefont{J.~R.}\ \bibnamefont{Andersen}}, \bibinfo
  {author} {\bibfnamefont{V.}~\bibnamefont{Del~Duca}}, \bibinfo {author}
  {\bibfnamefont{S.}~\bibnamefont{Frixione}}, \bibinfo {author}
  {\bibfnamefont{C.~R.}\ \bibnamefont{Schmidt}},\ and\ \bibinfo {author}
  {\bibfnamefont{W.~J.}\ \bibnamefont{Stirling}},\ }%
  \bibfield{journal}{%
  \bibinfo {journal} {JHEP}\ }%
  \textbf{\bibinfo {volume} {02}},\ \bibinfo {pages} {007} (\bibinfo {year}
  {2001}),\ \Eprint{http://arxiv.org/abs/hep-ph/0101180}{hep-ph/0101180}%
  \bibAnnoteFile{NoStop}{Andersen:2001kta}%
\bibitem{Fontannaz:2001nq}%
  \BibitemOpen
  \bibfield{author}{%
  \bibinfo {author} {\bibfnamefont{M.}~\bibnamefont{Fontannaz}}, \bibinfo
  {author} {\bibfnamefont{J.~P.}\ \bibnamefont{Guillet}},\ and\ \bibinfo
  {author} {\bibfnamefont{G.}~\bibnamefont{Heinrich}},\ }%
  \bibfield{journal}{%
  \bibinfo {journal} {Eur. Phys. J.}\ }%
  \textbf{\bibinfo {volume} {C22}},\ \bibinfo {pages} {303} (\bibinfo {year}
  {2001}),\ \Eprint{http://arxiv.org/abs/hep-ph/0107262}{hep-ph/0107262}%
  \bibAnnoteFile{NoStop}{Fontannaz:2001nq}%
\bibitem{Aurenche:2008dn}%
  \BibitemOpen
  \bibfield{author}{%
  \bibinfo {author} {\bibfnamefont{P.}~\bibnamefont{Aurenche}}, \bibinfo
  {author} {\bibfnamefont{R.}~\bibnamefont{Basu}},\ and\ \bibinfo {author}
  {\bibfnamefont{M.}~\bibnamefont{Fontannaz}},\ }%
  \bibfield{journal}{%
  \Doi{10.1140/epjc/s10052-008-0731-2}{\bibinfo {journal} {Eur. Phys. J.}}\ }%
  \textbf{\bibinfo {volume} {C57}},\ \bibinfo {pages} {681} (\bibinfo {year}
  {2008}),\ \Eprint{http://arxiv.org/abs/0807.2133}{arXiv:0807.2133 [hep-ph]}%
  \bibAnnoteFile{NoStop}{Aurenche:2008dn}%
\bibitem{Orr:1997im}%
  \BibitemOpen
  \bibfield{author}{%
  \bibinfo {author} {\bibfnamefont{L.~H.}\ \bibnamefont{Orr}}\ and\ \bibinfo
  {author} {\bibfnamefont{W.~J.}\ \bibnamefont{Stirling}},\ }%
  \bibfield{journal}{%
  \bibinfo {journal} {Phys. Rev.}\ }%
  \textbf{\bibinfo {volume} {D56}},\ \bibinfo {pages} {5875} (\bibinfo {year}
  {1997}),\ \Eprint{http://arxiv.org/abs/hep-ph/9706529}{hep-ph/9706529}%
  \bibAnnoteFile{NoStop}{Orr:1997im}%
\bibitem{DelDuca:1994ng}%
  \BibitemOpen
  \bibfield{author}{%
  \bibinfo {author} {\bibfnamefont{V.}~\bibnamefont{Del~Duca}}\ and\ \bibinfo
  {author} {\bibfnamefont{C.~R.}\ \bibnamefont{Schmidt}},\ }%
  \bibfield{journal}{%
  \Doi{10.1103/PhysRevD.51.2150}{\bibinfo {journal} {Phys. Rev.}}\ }%
  \textbf{\bibinfo {volume} {D51}},\ \bibinfo {pages} {2150} (\bibinfo {year}
  {1995}),\ \Eprint{http://arxiv.org/abs/hep-ph/9407359}{hep-ph/9407359}%
  \bibAnnoteFile{NoStop}{DelDuca:1994ng}%
\bibitem{Marquet:2007xx}%
  \BibitemOpen
  \bibfield{author}{%
  \bibinfo {author} {\bibfnamefont{C.}~\bibnamefont{Marquet}}\ and\ \bibinfo
  {author} {\bibfnamefont{C.}~\bibnamefont{Royon}},\ }%
  \bibfield{journal}{%
  \Doi{10.1103/PhysRevD.79.034028}{\bibinfo {journal} {Phys. Rev.}}\ }%
  \textbf{\bibinfo {volume} {D79}},\ \bibinfo {pages} {034028} (\bibinfo {year}
  {2009}),\ \Eprint{http://arxiv.org/abs/0704.3409}{arXiv:0704.3409 [hep-ph]}%
  \bibAnnoteFile{NoStop}{Marquet:2007xx}%
\bibitem{Brodsky:2011ig}%
  \BibitemOpen
  \bibfield{author}{%
  \bibinfo {author} {\bibfnamefont{S.~J.}\ \bibnamefont{Brodsky}}\ and\
  \bibinfo {author} {\bibfnamefont{L.}~\bibnamefont{Di~Giustino}},\ }%
  \bibfield{journal}{%
  \Doi{10.1103/PhysRevD.86.085026}{\bibinfo {journal} {Phys. Rev.}}\ }%
  \textbf{\bibinfo {volume} {D86}},\ \bibinfo {pages} {085026} (\bibinfo {year}
  {2012}),\ \Eprint{http://arxiv.org/abs/1107.0338}{arXiv:1107.0338 [hep-ph]}%
  \bibAnnoteFile{NoStop}{Brodsky:2011ig}%
\bibitem{Brodsky:2011ta}%
  \BibitemOpen
  \bibfield{author}{%
  \bibinfo {author} {\bibfnamefont{S.~J.}\ \bibnamefont{Brodsky}}\ and\
  \bibinfo {author} {\bibfnamefont{X.-G.}\ \bibnamefont{Wu}},\ }%
  \bibfield{journal}{%
  \Doi{10.1103/PhysRevD.85.034038, 10.1103/PhysRevD.86.079903}{\bibinfo
  {journal} {Phys. Rev.}}\ }%
  \textbf{\bibinfo {volume} {D85}},\ \bibinfo {pages} {034038} (\bibinfo {year}
  {2012}),\ \Eprint{http://arxiv.org/abs/1111.6175}{arXiv:1111.6175 [hep-ph]}%
  \bibAnnoteFile{NoStop}{Brodsky:2011ta}%
\bibitem{Brodsky:2012rj}%
  \BibitemOpen
  \bibfield{author}{%
  \bibinfo {author} {\bibfnamefont{S.~J.}\ \bibnamefont{Brodsky}}\ and\
  \bibinfo {author} {\bibfnamefont{X.-G.}\ \bibnamefont{Wu}},\ }%
  \bibfield{journal}{%
  \Doi{10.1103/PhysRevLett.109.042002}{\bibinfo {journal} {Phys. Rev. Lett.}}\
  }%
  \textbf{\bibinfo {volume} {109}},\ \bibinfo {pages} {042002} (\bibinfo {year}
  {2012}),\ \Eprint{http://arxiv.org/abs/1203.5312}{arXiv:1203.5312 [hep-ph]}%
  \bibAnnoteFile{NoStop}{Brodsky:2012rj}%
\bibitem{Brodsky:2012ik}%
  \BibitemOpen
  \bibfield{author}{%
  \bibinfo {author} {\bibfnamefont{S.~J.}\ \bibnamefont{Brodsky}}\ and\
  \bibinfo {author} {\bibfnamefont{X.-G.}\ \bibnamefont{Wu}},\ }%
  \bibfield{journal}{%
  \Doi{10.1103/PhysRevD.85.114040}{\bibinfo {journal} {Phys. Rev.}}\ }%
  \textbf{\bibinfo {volume} {D85}},\ \bibinfo {pages} {114040} (\bibinfo {year}
  {2012}),\ \Eprint{http://arxiv.org/abs/1205.1232}{arXiv:1205.1232 [hep-ph]}%
  \bibAnnoteFile{NoStop}{Brodsky:2012ik}%
\bibitem{Mojaza:2012mf}%
  \BibitemOpen
  \bibfield{author}{%
  \bibinfo {author} {\bibfnamefont{M.}~\bibnamefont{Mojaza}}, \bibinfo {author}
  {\bibfnamefont{S.~J.}\ \bibnamefont{Brodsky}},\ and\ \bibinfo {author}
  {\bibfnamefont{X.-G.}\ \bibnamefont{Wu}},\ }%
  \bibfield{journal}{%
  \Doi{10.1103/PhysRevLett.110.192001}{\bibinfo {journal} {Phys. Rev. Lett.}}\
  }%
  \textbf{\bibinfo {volume} {110}},\ \bibinfo {pages} {192001} (\bibinfo {year}
  {2013}),\ \Eprint{http://arxiv.org/abs/1212.0049}{arXiv:1212.0049 [hep-ph]}%
  \bibAnnoteFile{NoStop}{Mojaza:2012mf}%
\bibitem{Wu:2013ei}%
  \BibitemOpen
  \bibfield{author}{%
  \bibinfo {author} {\bibfnamefont{X.-G.}\ \bibnamefont{Wu}}, \bibinfo {author}
  {\bibfnamefont{S.~J.}\ \bibnamefont{Brodsky}},\ and\ \bibinfo {author}
  {\bibfnamefont{M.}~\bibnamefont{Mojaza}},\ }%
  \bibfield{journal}{%
  \Doi{10.1016/j.ppnp.2013.06.001}{\bibinfo {journal} {Prog. Part. Nucl.
  Phys.}}\ }%
  \textbf{\bibinfo {volume} {72}},\ \bibinfo {pages} {44} (\bibinfo {year}
  {2013}),\ \Eprint{http://arxiv.org/abs/1302.0599}{arXiv:1302.0599 [hep-ph]}%
  \bibAnnoteFile{NoStop}{Wu:2013ei}%
\bibitem{Brodsky:2013vpa}%
  \BibitemOpen
  \bibfield{author}{%
  \bibinfo {author} {\bibfnamefont{S.~J.}\ \bibnamefont{Brodsky}}, \bibinfo
  {author} {\bibfnamefont{M.}~\bibnamefont{Mojaza}},\ and\ \bibinfo {author}
  {\bibfnamefont{X.-G.}\ \bibnamefont{Wu}}}%
   (\bibinfo {year} {2013}),\
  \Eprint{http://arxiv.org/abs/1304.4631}{arXiv:1304.4631 [hep-ph]}%
  \bibAnnoteFile{NoStop}{Brodsky:2013vpa}%
\bibitem{Zheng:2013uja}%
  \BibitemOpen
  \bibfield{author}{%
  \bibinfo {author} {\bibfnamefont{X.-C.}\ \bibnamefont{Zheng}}, \bibinfo
  {author} {\bibfnamefont{X.-G.}\ \bibnamefont{Wu}}, \bibinfo {author}
  {\bibfnamefont{S.-Q.}\ \bibnamefont{Wang}}, \bibinfo {author}
  {\bibfnamefont{J.-M.}\ \bibnamefont{Shen}},\ and\ \bibinfo {author}
  {\bibfnamefont{Q.-L.}\ \bibnamefont{Zhang}}\ }%
  \textbf{\bibinfo {volume} {JHEP10}},\ \bibinfo {pages} {117} (\bibinfo {year}
  {2013}),\ \Eprint{http://arxiv.org/abs/1308.2381}{arXiv:1308.2381 [hep-ph]}%
  \bibAnnoteFile{NoStop}{Zheng:2013uja}%
\end{thebibliography}
\end{document}